\RequirePackage{fix-cm}
\documentclass[twocolumn]{svjour3}          
\smartqed  
\usepackage{graphicx}
\usepackage{bm}
\usepackage{epsfig}
\usepackage{amsmath}
\usepackage{graphicx,color}
\usepackage{comment}
\begin{document}

\title{General Relativistic effects in the structure of massive white dwarfs
}


\author{G. A. Carvalho         \and
  R. M. Marinho Jr \and
  M. Malheiro 
}


\institute{
              $^1$Department of Physics, Technological Institute of Aeronautics Pra\c{c}a Marechal Eduardo Gomes 50, Sao Jose dos Campos, SP, Brazil\\
              \email{araujogc@ita.br}           
}

\date{Received: date / Accepted: date}

\maketitle

\begin{abstract} 
In this work we investigate the structure of white dwarfs using the Tolman-Oppenheimer-Volkoff equations and compare our results with those obtained from Newtonian equations of gravitation in order to put in evidence the importance of General Relativity (GR) for the structure of such stars. We consider in this work for the matter inside white dwarfs two equations of state, frequently found in the literature, namely, the Chandrasekhar and Salpeter equations of state. We find that using Newtonian equilibrium equations, the radii of massive white dwarfs ($M>1.3M_{\odot}$) are overestimated in comparison with GR outcomes. For a mass of $1.415M_{\odot}$ the white dwarf radius predicted by GR is about 33\% smaller than the Newtonian one. Hence, in this case, for the surface gravity the difference between the general relativistic and Newtonian outcomes is about 65\%. We depict the general relativistic mass-radius diagrams as $M/M_{\odot}=R/(a+bR+cR^2+dR^3+kR^4)$, where $a$, $b$, $c$ and $d$ are parameters obtained from a fitting procedure of the numerical results and $k=(2.08\times 10^{-6}R_{\odot})^{-1}$, being $R_{\odot}$ the radius of the Sun in km. Lastly, we point out that GR plays an important role to determine any physical quantity that depends, simultaneously, on the mass and radius of massive white dwarfs.

\keywords{First keyword \and Second keyword \and More}
\end{abstract}

\section{Introduction}
\label{intro}

Massive white dwarfs (WD) were documented in \cite{Kepler2007,kepler2013,Haensel2004,Vennes1997} reaching for the first time values nearby to the Chandrasekhar mass limit of $1.44M_{\odot}$ \cite{Chandrasekhar1935}, with $M_{\odot}$ being the mass of the Sun. Particularly, in the Extreme Ultraviolet Explorer all-sky survey (EUVE) one can find several data of massive WD. The most massive WD presented in the EUVE observations have a mass of $1.41M_{\odot}$ \cite{Vennes1997}. Also in Refs. \cite{kepler2013}-\cite{Haensel2004} a considerable amount of WD with masses between $1.32-1.37M_{\odot}$ can be found.

Moreover, recent observations reveals the existence of some super-luminous type Ia supernovae, e.g, SN200-6gz, SN2007if, SN2009dc \cite{hicken2007,yamanaka2009,scalzo2010,taub2011} (for a review about this kind of astrophysical event we quote \cite{Leibundgut2008}). Some authors suggest that possible explanations for these objects are super-Chandrasekhar WD \cite{howell2006,scalzo2010,silverman2010}. To achieve super-Chandrasekhar WD authors of Refs. \cite{Frazon2015,Otoniel2017,Das2014,Das2015,Bera2016a} consider the WD in a presence of very strong magnetic fields. However, those putative WD with strong magnetic fields were showed to be unstable in \cite{Coelho2013}.
Also some attention has been driven to the description of some anomalous X-ray pulsars (AXP) and soft gamma-ray repeaters (SGR) as highly magnetized very massive rotation powered white dwarf pulsars \cite{lobato,Malheiro2012,Rueda2013,Jaziel2014,Rao2016,Jaziel2017}. In view of these new discoveries it is worth to better identify how General Relativity (GR) can affect the structure of very massive WD.

Indeed, since the discovery of WD, the discussion about the importance of GR for these objects was established. The general relativistic effects on the structure of WD was first qualitatively discussed by Kaplan \cite{kaplan}. Fifteen years later Chandrasekhar derived the instability criteria in a general relativistic framework \cite{chandratooper}. Kaplan concluded that when the WD radius becomes smaller than $1.1\times 10^3 ~{\rm km}$, GR would probably induces a dynamical instability in the star. Furthermore, Chandrasekhar \cite{chandratooper} concluded that general relativistic effects leads to a smaller critical central density and, consequently, it limits the value of the radius. In addition, J. Cohen {\it et al} \cite{cohen} studied the oscillation period of WD taking GR effects into account and they argue that general relativistic WD unlike Newtonian ones have a minimum fundamental period for a given composition.

In more recent works \cite{rueda,rueda2,rueda3} calculations were made for general relativistic uniformly-rotating WD. Summarizing those results: they showed that the rotation can have astrophysical implications to Soft-Gamma Repeaters and Anomalous X-Ray Pulsars, rotation can also uplift slightly the maximum stable mass configuration and they also showed that the spin-up and spin-down eras are different for sub and super-Chandrasekhar WD. In addition, in \cite{boshkayev2015} the authors compare the results of general relativistic uniformly-rotating WD with uniformly-rotating Newtonian WD. In light of uniformly-rotating case the minimum rotation period is about the same in both cases (general relativistic and Newtonian) and the general relativistic maximum mass is slightly below the Newtonian maximum mass. However, they also showed by the turning point method that general relativistic WD can be axis-symmetrically unstable while Newtonian WD are stable.

Turning to the static case, our main aim in the present work is to show that general relativistic effects are relevant for the determination of the radius of massive WD.
In fact, there are some works where general relativistic calculations of the mass-radius relation were made for a static model of WD \cite{irina,mathewnandy2014,Bera2016,Rotondo2011a,Rotondo2011b}. In those works authors find that general relativistic hydrostatic equilibrium yields to a maximum mass slightly below the Chandrasekhar limiting mass. Nevertheless, the role of GR for the radius of very massive WD ($M>1.3M_{\odot}$) was not stressed. Thus, in the next sections of this paper we address, minutely, the general relativistic effects on the mass-radius relation of massive WD, showing that GR is quite important to determine the radii of those massive WD. Moreover, considering the observational data obtained for very massive WD, GR turns out to be very relevant to estimate, precisely, the WD radius and others WD properties, such as surface gravity.

\section{Equation of state}
The result of Chandrasekhar mass limit is one of the most established astrophysical constraints since there is no confirmed observational data of WD with masses above $1.44M_{\odot}$, until now. Therefore, in the present work we use the Chandrasekhar equation of state, i.e., the equation of state that describes a fully degenerate relativistic electron gas. In such a model the pressure and total energy density of the fluid are given, respectively, by \cite{chandra,Rey2008}
  \begin{equation}\label{eos}
  p(k_F) = \frac{1}{3\pi^2\hbar^3}\int_0^{k_F}\frac{k^4c^2}{\sqrt{k^2c^2+m_e^2c^4}}dk,
  \end{equation}
  \begin{eqnarray}\label{energy} 
    \epsilon(k_F) &=& \rho c^2 + \epsilon_e\\
    &=& \frac{m_N\mu_ek_F^3}{3\pi^2 \hbar^3}c^2 + \frac{1}{\pi^2\hbar^3}\int_0^{k_F}\sqrt{k^2c^2+m_e^2c^4}k^2dk \nonumber,
  \end{eqnarray}
where $c$ is the speed of light, $m_e$ is the electron mass, $m_N$ the nucleon mass, $\hbar$ the reduced Planck constant, $\mu_e$ is the ratio between the nucleon number $A$ and atomic number $Z$ for ions and $k_F$ is the Fermi momentum of the electron. Eq. \eqref{eos} is the isotropic electron degeneracy pressure. The first and second terms in Eq. \eqref{energy} are, respectively, the energy density related to the rest mass of the ions and the electron energy density.\\
Eqs. \eqref{eos} and \eqref{energy} can be put in a simpler form in order to favor future numerical calculations    
\begin{subequations}\label{eos:simp}
\begin{align}
p(x) &= \epsilon_0 f(x), \\
\epsilon(x) &= \epsilon_0 g(x),
\end{align}
\end{subequations}
where $\epsilon_0=m_ec^2/\pi^2\lambda_e^3$, with $\lambda_e$ being the electron Compton wavelength, $x=k_F/m_ec$ the dimensionless Fermi momentum and the functions $f(x)$ and $g(x)$ using $\mu_e=\frac{A}{Z}=2$ are
\begin{equation}
f(x)=\frac{1}{24}\left[(2x^3-3x)\sqrt{x^2+1}+3\textrm{asinh} x\right],
\end{equation}
\begin{equation}
g(x)=1215.26x^3+\frac{1}{8}\left[(2x^3+x)\sqrt{x^2+1}-\textrm{asinh} x\right].
\end{equation}
In terms of the dimensionless Fermi momentum $x$ the mass density becomes
\begin{equation}
\rho=9.738\times  10^5 \mu_e x^3 ~{\rm g/cm^3}.
\end{equation}

Salpeter in a seminal paper \cite{Salpeter} have improved the above EoS by considering several corrections, such as: electrostatic corrections due to Coulomb interaction, deviations of the electron charge distribution from uniformity, inverse beta-decay process, inclusion of correlation and exchange energies. Hamada \& Salpeter \cite{Hamada1961} have showed, {\it a posteriori}, that the mass-radius relation of WD are modified in a nontrivial way depending on the interior composition due to those corrections applied by Salpeter.

\section{Hydrostatic equilibrium equations}\label{hydro}

\subsection{Newtonian case}\label{newtoncase}
We assume in our calculations that the mass configuration of the star is static and spherically symmetric. In this case the pressure and the density of the fluid are functions of the radial coordinate $r$ only. Hence, for the structure of a Newtonian star we have the following equilibrium equations \cite{chandra}
\begin{subequations}\label{newtonian1}
\begin{align}
\frac{dp}{dr}&=-\frac{Gm\rho}{r^2},\label{first}\\\nonumber\\
\frac{dm}{dr}&=4\pi r^2\rho,\label{second}
\end{align}
\end{subequations}
where $p$ denotes pressure, $G$ is the gravitational constant, $\rho$ is the mass density, and $m$ represents the enclosed mass inside a sphere of radius $r$.

\subsection{Special relativistic case}\label{newtoncase2} 
In addition to Newtonian case one may consider special relativistic improvements. For such, the mass density is replaced by the total energy density of the system (see Eq. \eqref{energy}), consequently, the relativistic kinetic energy of the electrons are taken into account, such that Eqs. \eqref{first} and \eqref{second} becomes
\begin{subequations}\label{newtonian2}
\begin{align}
\frac{dp}{dr}&=-\frac{Gm(\rho c^2+e_e)}{r^2c^2}=-\frac{Gm\epsilon}{r^2c^2},\label{first2}\\\nonumber\\
\frac{dm}{dr}&=\frac{4\pi r^2(\rho c^2+e_e)}{c^2}=\frac{4\pi r^2\epsilon}{c^2}.\label{second2}
\end{align}
\end{subequations}
Along the present paper we will refer to this case as special relativistic (SR) case.

\subsection{General relativistic case}\label{tovcase}
To derive the hydrostatic equilibrium equations in a general relativistic framework it is used the interior Schwarzschild solution and the energy-momentum tensor of a perfect fluid \cite{adler}. For a detailed derivation of the general relativistic hydrostatic equilibrium equation see \cite{Oppenheimer1939,Tolman1939,Gautreau1990}.
The Eqs. \eqref{first} and \eqref{second} now reads
\begin{subequations}\label{tov.}
\begin{align}
\frac{dp}{dr}&=-(\epsilon+p) \frac{d\phi}{dr}\nonumber\\ &=-\frac{Gm\epsilon}{c^2r^2}\left[1+\frac{p}{\epsilon}\right]\left[1+\frac{4\pi r^3p}{mc^2}\right]\left[1-\frac{2Gm}{c^2r}\right]^{-1},\label{first1}\\\nonumber\\
\frac{dm}{dr}&=\frac{4\pi r^2\epsilon}{c^2}\label{second1},
\end{align}
\end{subequations}
being $e^{2\phi}$ the temporal metric coefficient $g_{00}$. In the weak field limit $g_{00}=1+2\Phi/c^2+\mathcal{O}^2$, where $\Phi$ corresponds to the Newtonian gravitational potential. So, the formal definition of the gravitational field of a static and spherically symmetric object in GR corresponds to $g_{GR}=-d\phi/dr$, where $\phi$ represents the general relativistic gravitational potential. 

After integrating the above equations the interior Schwarzschild solution is matched smoothly with the vacuum exterior Schwarzschild line element.

The new three terms in square brackets of the equilibrium equation \eqref{first1} are general relativistic corrections terms. From Eq. \eqref{first1} is reliable that the general relativistic effects become relevant when the star is sufficiently compact, i.e., when the factor $2Gm/c^2r$ approaches unity and when the pressure is high enough to become comparable to the energy density of the fluid, i.e., when $p/\epsilon$ and $4\pi r^3p/mc^2$ are comparable with the unity.  


\section{Initial and boundary conditions}
In this paper we use the equation of state \eqref{eos:simp} to solve the equilibrium equations through a forth-order Runge-Kutta method. The initial conditions are
\begin{equation}
p(r=0)=p_c,\quad \rho(r=0)=\rho_c \quad {\rm and} \quad m(r=0)=0. 
\end{equation}
The star's surface is reached when the pressure vanishes, consequently the energy density (or mass density) also goes to zero at the surface. Therefore the boundary conditions reads
\begin{equation}
p(r=R)= 0, \quad \rho(r=R)=0 \quad {\rm and} \quad m(r=R)=M, 
\end{equation}
where $R$ and $M$ mean the total radius and total mass of the star, respectively. Since one may use several values for the central pressure $p_c$ a family of solutions can be found for the star mass and radius.

\section{Comparison between Newtonian and general relativistic cases}\label{comparison}

\begin{figure}[h]
\begin{center}
\includegraphics[width=0.8\linewidth]{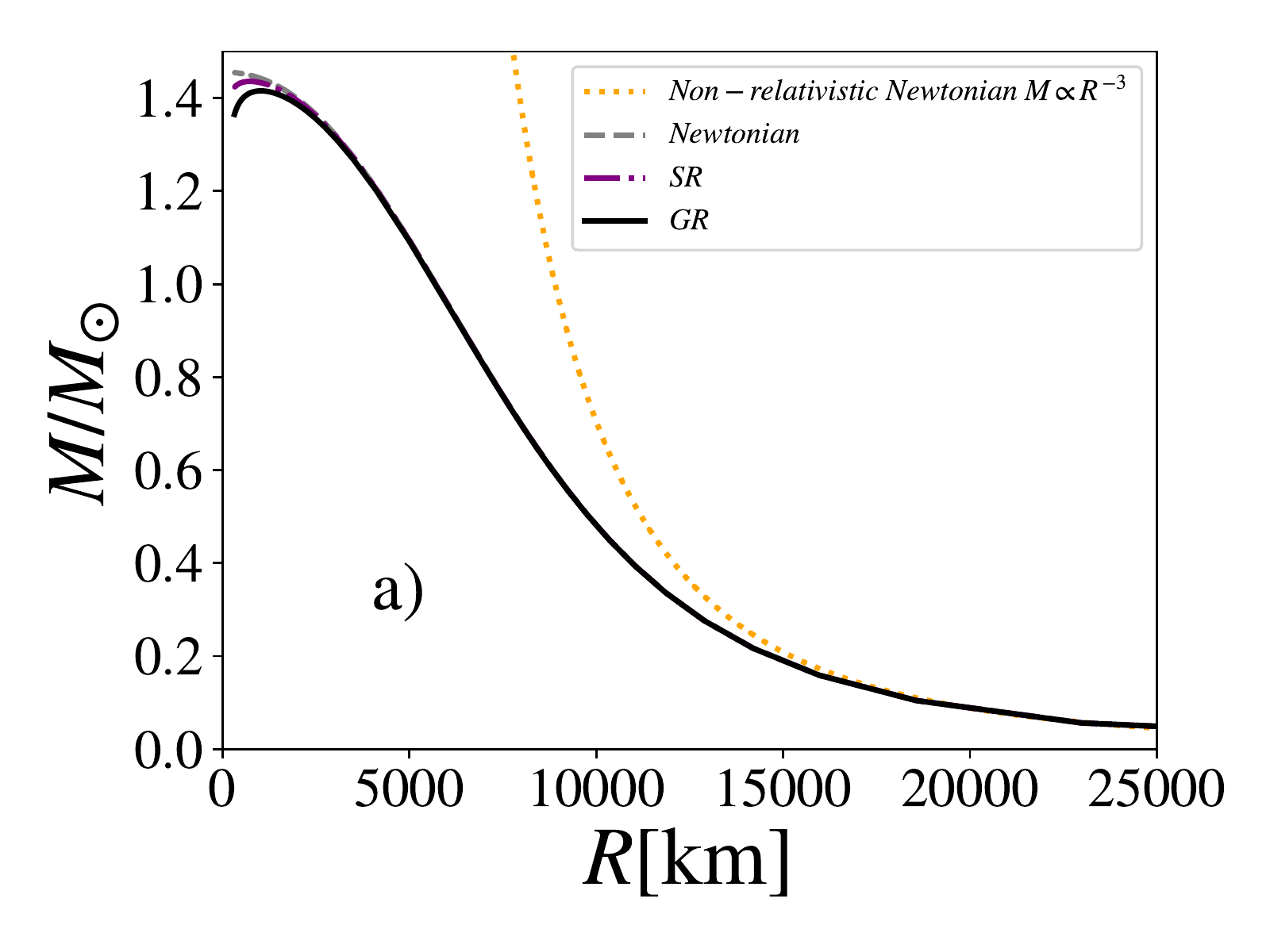}
\includegraphics[width=0.8\linewidth]{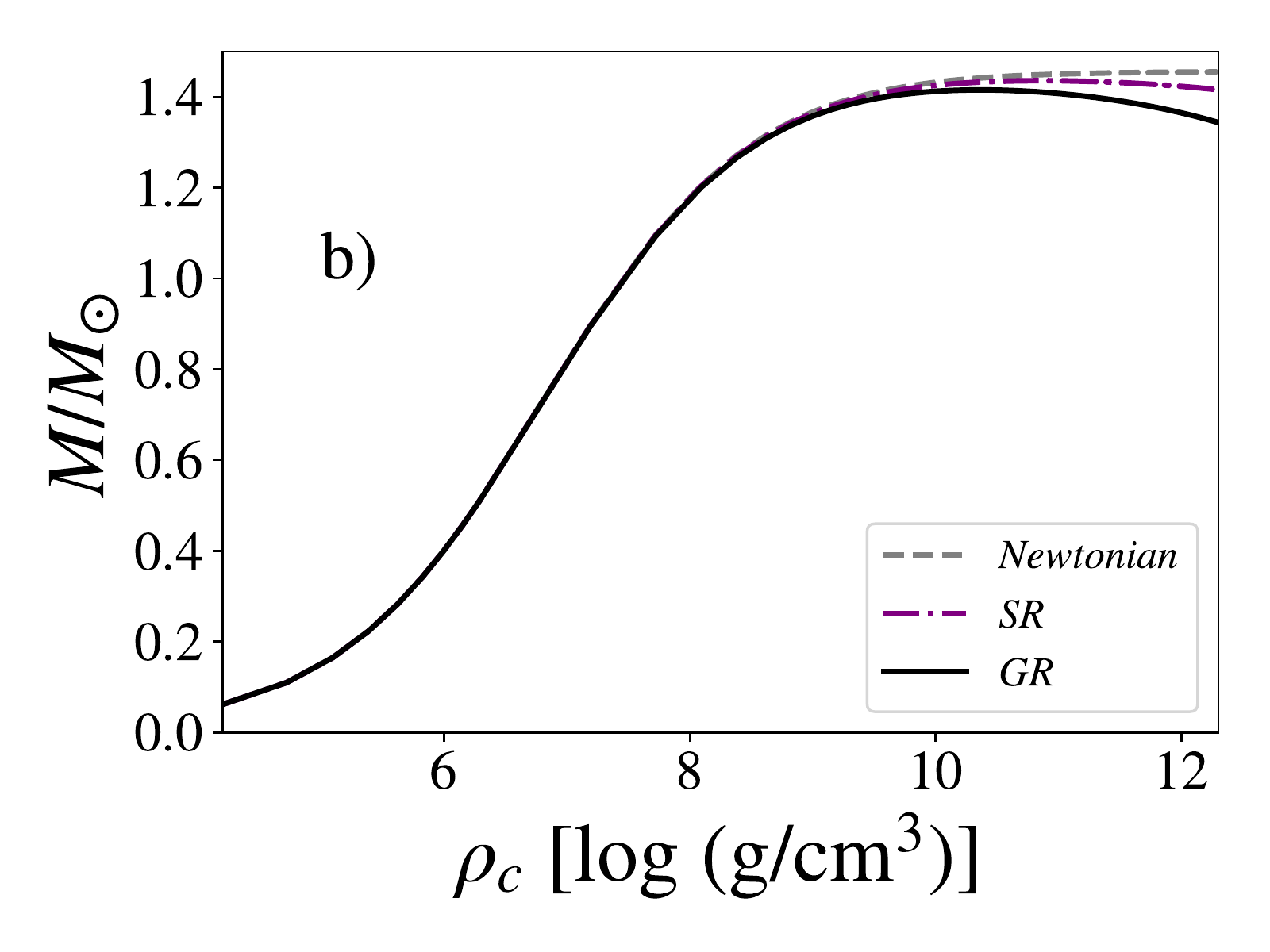}
\caption{a) Mass-radius relation and b) mass-central mass density relation for general relativistic and Newtonian cases. In both plots the solid black line represents the outcomes for general relativistic WD, the dashed-dotted magenta line represents Newtonian results with special relativistic corrections and the dashed gray line represents the Newtonian results. It is also displayed in a) the non-relativistic mass-radius relation (dotted orange line), in which $M\propto 1/R^3$.}
\label{mrgeral.}
\end{center}
\end{figure}

\begin{figure}[h]
\begin{center}
\includegraphics[width=0.8\linewidth]{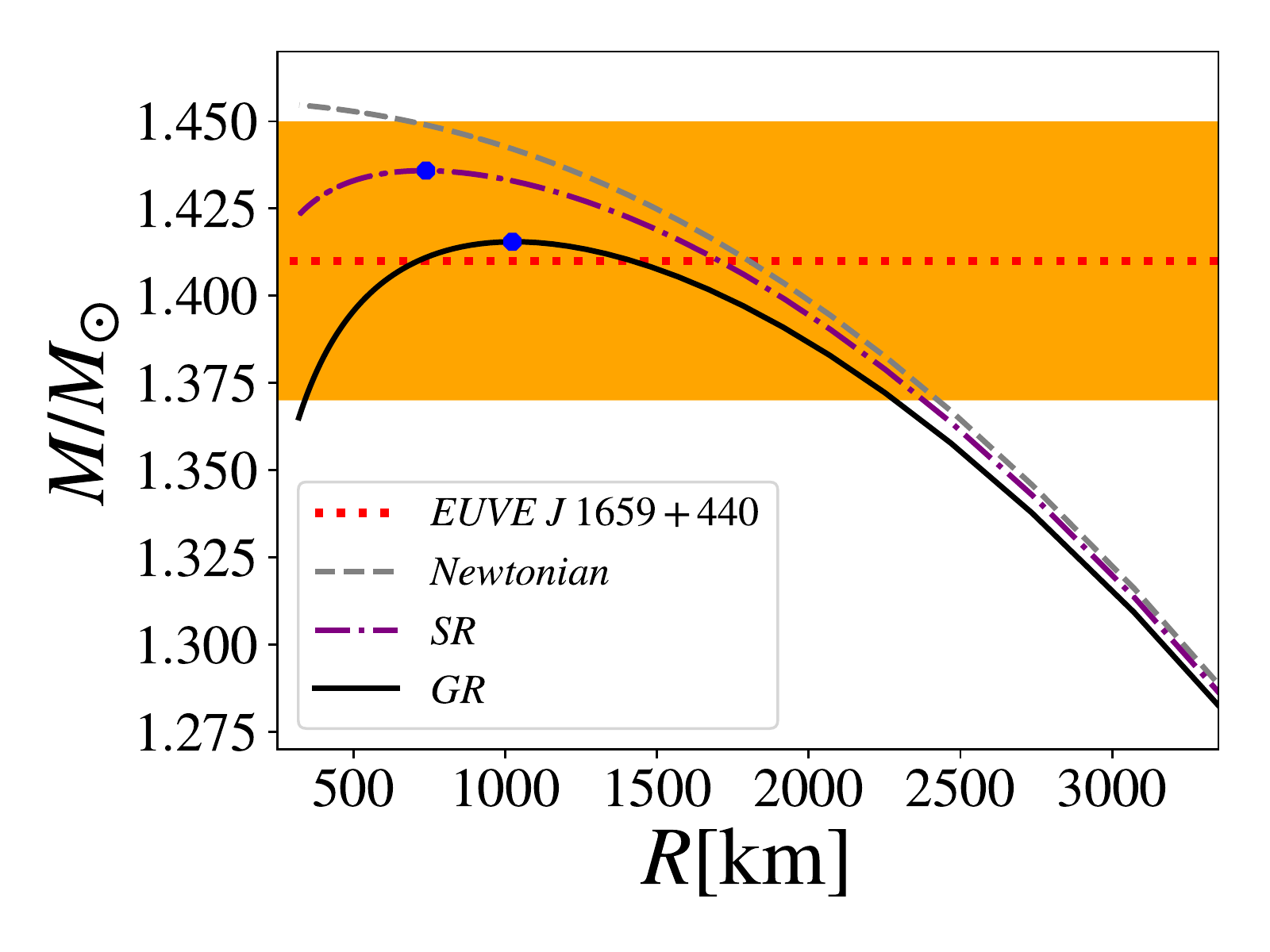}
\caption{Mass-radius relation of massive WD. The curves follow the same representation as in Fig.\eqref{mrgeral.}. The full blue circles mark the maximum masses. The dotted red line represents the measured mass of the most massive white dwarf ($M=1.41M_{\odot}\pm 0.04$) found in literature \cite{Vennes1997} and the shaded orange region corresponds to its estimated error.}
\label{zoom}
\end{center}
\end{figure}

Using several values of central pressure $p_c$ we construct the mass-radius and mass-central density relations for the three cases previously explained.
From Fig.\eqref{mrgeral.} it can be seen that the purely Newtonian case (Sect. \eqref{newtoncase}) does not have the secular instability $\partial M/\partial R>0$ (more details, see \cite{chandratooper,Herrera2003,Penrose2002,Knutsen1988}), while SR case presents instability when the electrons are highly relativistic. This aspect is easier to see in Fig.\eqref{zoom}, where we highlight the region of massive WD for the mass-radius relation. We also display in Fig.\eqref{zoom} the observational data of the most massive white dwarf ($M=1.41M_{\odot}\pm 0.04$) found in literature \cite{Vennes1997}. 
\begin{table*}[t]
          \caption{Maximum mass and minimum radius for the static models of WD stars.}\label{tabela}
		\begin{center}
			\begin{tabular}{l l l}
				\hline
				Model	&	Mass/$M_{\odot}$				&	Radius(km)\\
				\hline
				Newtonian $P=P(\rho)$&	$1.4546$	&	$329$\\
				Special Relativity (SR) $P=P(\epsilon)$&	 $1.4358$&$739$\\
				General Relativity (GR) $P=P(\epsilon)$& $1.4154$&$1021$\\
				Non-relativistic Newtonian $P=K\rho^{5/3}$ & $1.4564$ & $7833$\\
				\hline
			\end{tabular}
		\end{center}
\end{table*}

From Fig.\eqref{zoom} it is worth to note that GR does not affect greatly the maximum mass, rather it diminishes the maximum stable mass a few percents $\sim 3\%$ (see also Tab.\eqref{tabela}). However, it is worthwhile to cite that the minimum radii, i.e., the radii corresponding to the predicted maximum masses, are very different. For instance, the minimum radius predicted by general relativistic calculations is about three times larger than Newtonian ones (see Tab.\eqref{tabela}). Similar results can be found in \cite{eu1,eu4}.

\subsection{Fixed total star mass}

\begin{figure}[h!]
\begin{center}
\includegraphics[width=0.8\linewidth]{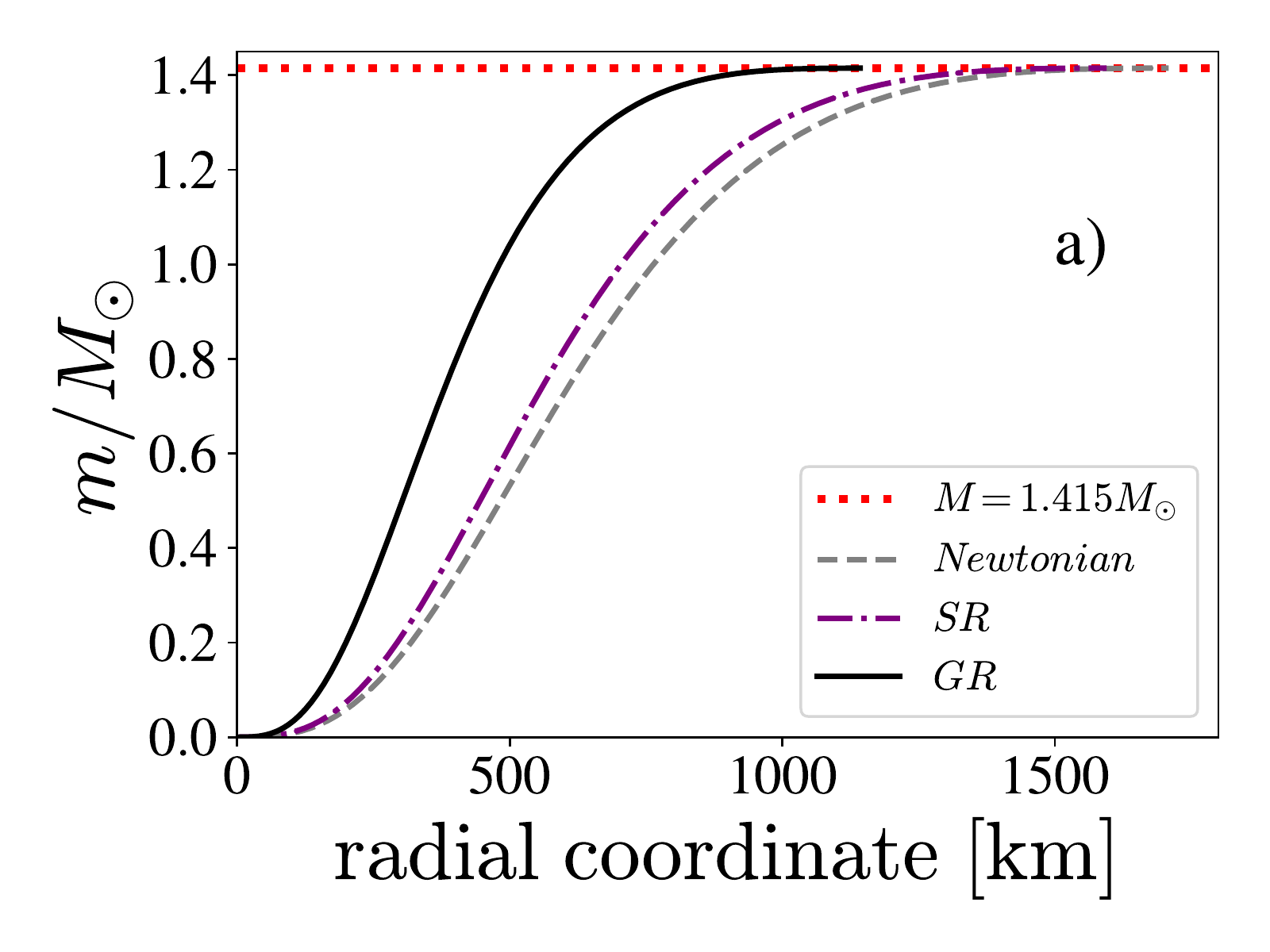}
\includegraphics[width=0.8\linewidth]{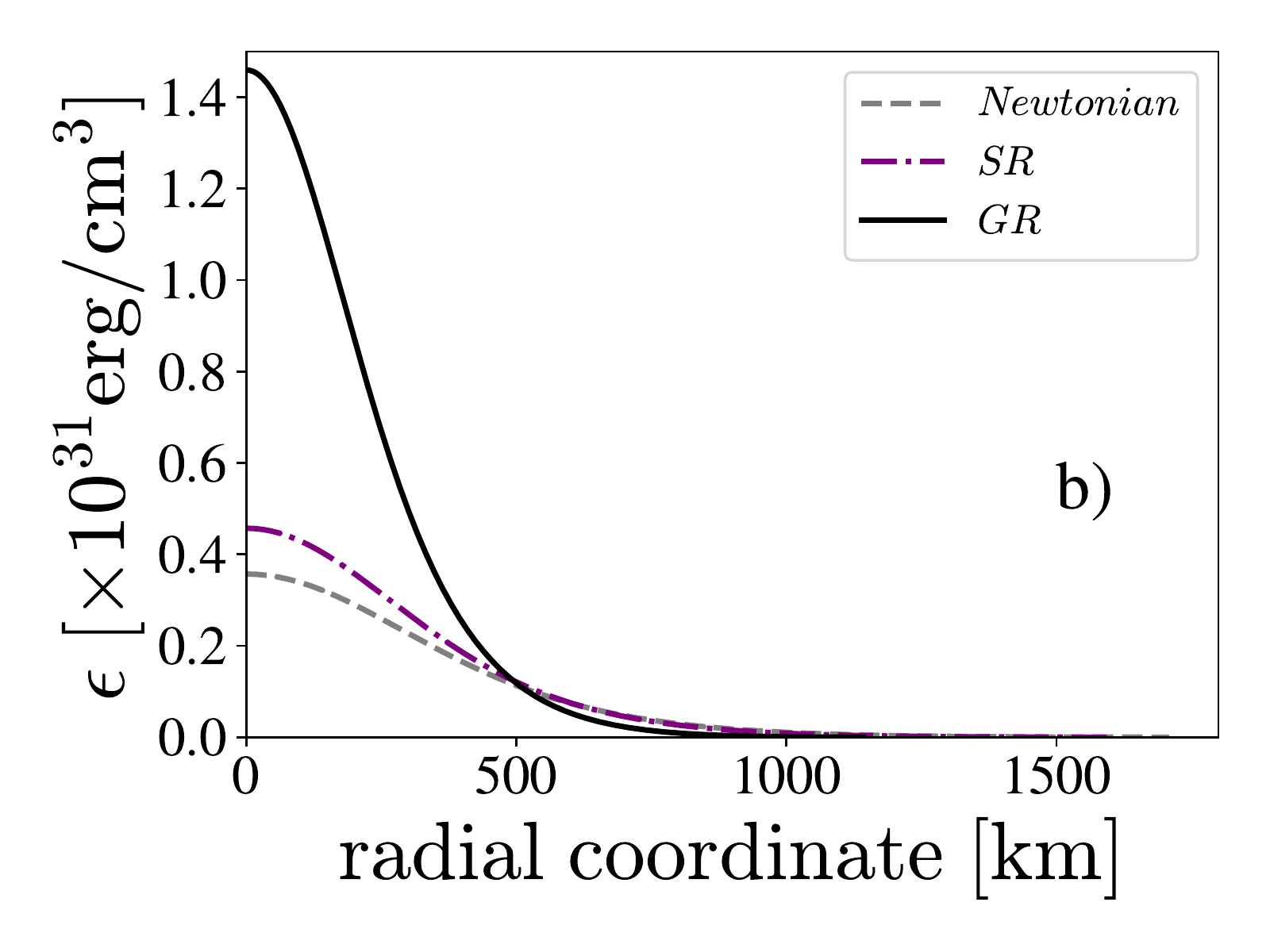}
\includegraphics[width=0.8\linewidth]{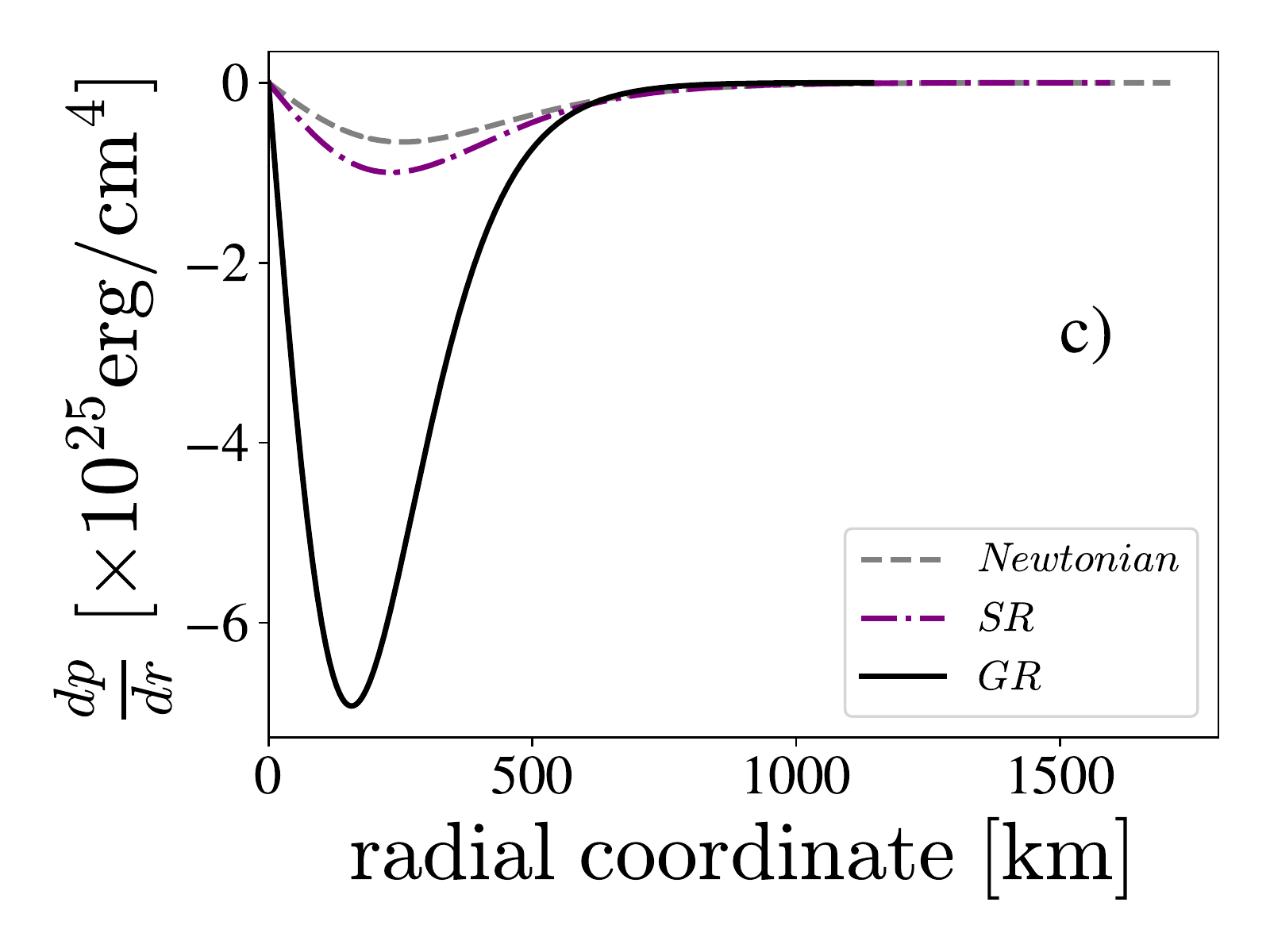}
\caption{From top to bottom: a) mass profiles, b) energy density profiles and c) gradient of pressure profiles. All profiles correspond to a fixed total star mass of $M=1.415M_{\odot}$.}
\label{varios}
\end{center}
\end{figure}

From Fig.\eqref{zoom} it can also be seen that for a fixed total star mass between $1.3-1.415M_{\odot}$ the values of radii are very sensitive depending on the case. Tab.\eqref{tabelaraios} shows the calculated radii for several values of total mass from Newtonian and general relativistic cases.
	\begin{table*}[t]
          \caption{Corresponding radii to fixed total star masses in Newtonian and general relativistic cases. $R_{\textrm{Newton}}$ means the radius predicted by Newtonian case (see Sect.\eqref{newtoncase}), $R_{\textrm{SR}}$ is the radius given by Sect.\eqref{newtoncase2}, $R_{\textrm{GR}}$ is the radius in general relativistic case given in Sect.\eqref{tovcase} and the $R_{\textrm{NR}}$ is the radius supplied by non-relativistic approximation, where the mass follows the relation $M/M_{\odot} \propto 1/R^3$.}\label{tabelaraios}
		\begin{center}
			\begin{tabular}{c c c c c}
				\hline
				Mass/$M_{\odot}$ 	&	$R_{\textrm{Newton}}$(km)    &    $R_{\textrm{SR}}$(km)   &   $R_{\textrm{GR}}$(km)   &   $R_{\textrm{NR}}$(km)\\
				\hline
				$1.300$ &	 $3241$	&	$3222$  &  $3185$  &  $8140$\\
                                $1.312$ &	 $3107$	&	$3081$  &  $3030$  &  $8114$\\
                                $1.325$ &	 $2969$	&	$2937$  &  $2878$  &  $8087$\\
                                $1.338$ &	 $2823$	&	$2788$  &  $2724$  &  $8062$\\
                                $1.351$ &	 $2671$	&	$2633$  &  $2562$  &  $8036$\\
                                $1.364$ &	 $2509$	&	$2467$  &  $2384$  &  $8011$\\
                                $1.376$ &	 $2336$	&	$2286$  &  $2179$  &  $7986$\\
                                $1.389$ &	 $2148$	&	$2085$  &  $1942$  &  $7961$\\
                                $1.402$ &	 $1942$	&	$1859$  &  $1656$  &  $7937$\\
                                $1.415$ &	 $1708$	&	$1595$  &  $1145$  &  $7913$\\
				\hline
			\end{tabular}
		\end{center}
	\end{table*}
Tab.\eqref{tabeladensidades} presents the values of central mass density for some values of fixed total mass for several cases. 
	\begin{table*}[t]
          \caption{Corresponding central mass densities to fixed total masses in Newtonian and general relativistic cases. $\rho_C^{\textrm{Newton}}$ means the central density achieved in Newtonian case (Sect. \eqref{newtoncase}), $\rho_C^{\textrm{SR}}$ is the central density given by SR case Sect.\eqref{newtoncase2}, $\rho_C^{\textrm{GR}}$ is the central density found for the general relativistic case Sect.\eqref{tovcase}.}\label{tabeladensidades}
		\begin{center}
			\begin{tabular}{c c c c}
				\hline
				Mass/$M_{\odot}$ 	&	$\rho_C^{\textrm{Newton}}$(g/cm$^3$)    &    $\rho_C^{\textrm{SR}}$(g/cm$^3$)   &   $\rho_C^{\textrm{GR}}$(g/cm$^3$)\\
				\hline
                                $1.376$ &	 $3.76\times 10^8$	&	$9.85\times 10^8$  &  $1.91\times 10^9$  \\
                                $1.389$ &	 $2\times 10^9$	        &	$2.05\times 10^9$  &  $2.28\times 10^9$  \\
                                $1.402$ &	 $2.28\times 10^9$	&	$2.82\times 10^9$  &  $4.51\times 10^9$  \\
                                $1.415$ &	 $4.08\times 10^9$	&	$5.05\times 10^9$  &  $1.61\times 10^{10}$  \\
				\hline
			\end{tabular}
		\end{center}
	\end{table*}
In Fig.\eqref{varios} we show, for a fixed total mass of $M=1.415M_{\odot}$, the profiles of mass, gradient of pressure and energy density for the three cases. We remark from Fig.\eqref{varios}, that in order to obtain the same total mass in all cases the structure of the stars are very distinct. From Fig.\eqref{varios} we can note that in general relativistic case the energy density in the central region of the star is larger than in Newtonian cases. This effect at same time makes the WD's mass more concentrated at the star center and the pressure gradient to decay more sharply for the general relativistic calculations. 
One can calculate the Newtonian gravitational field as 
\begin{equation}
g_{\textrm{Newton}}=-\frac{Gm}{r^2},
\end{equation}
and the general relativistic gravitational field from \eqref{first1} becomes
\begin{equation}\label{surfgravityGR}
g_{\textrm{GR}}=-\frac{Gm}{r^2}\left[1+\frac{4\pi r^3p}{mc^2}\right]\left[1-\frac{2Gm}{c^2r}\right]^{-1}.
\end{equation}

\begin{figure}[h]
\begin{center}
\includegraphics[width=0.8\linewidth]{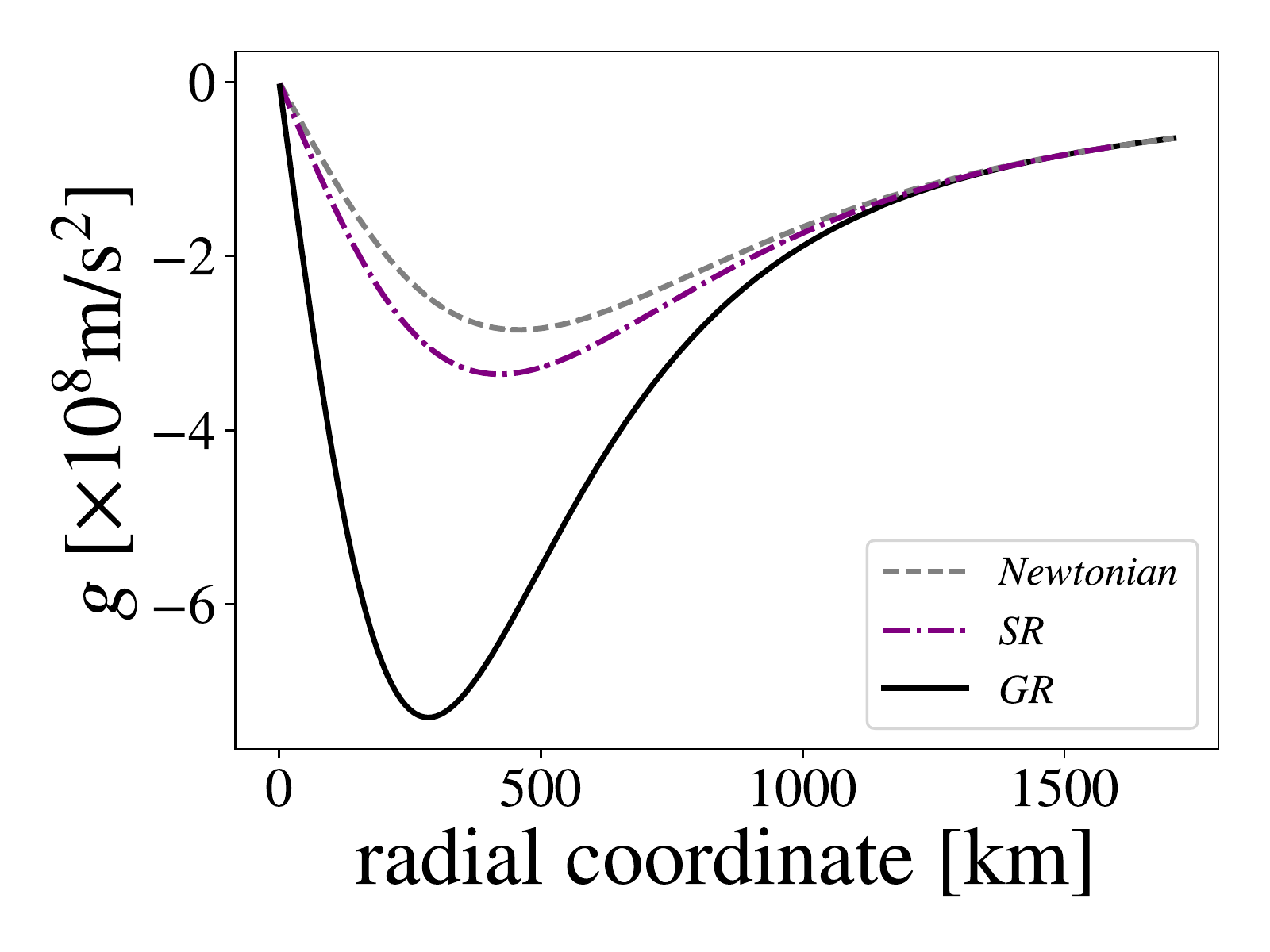}
\caption{General relativistic and Newtonian gravitational fields as a function of radial coordinate for a fixed total star mass of $1.415M_{\odot}$.}
\label{potentials}
\end{center}
\end{figure}

\begin{figure}[h]
\begin{center}
\includegraphics[width=0.8\linewidth]{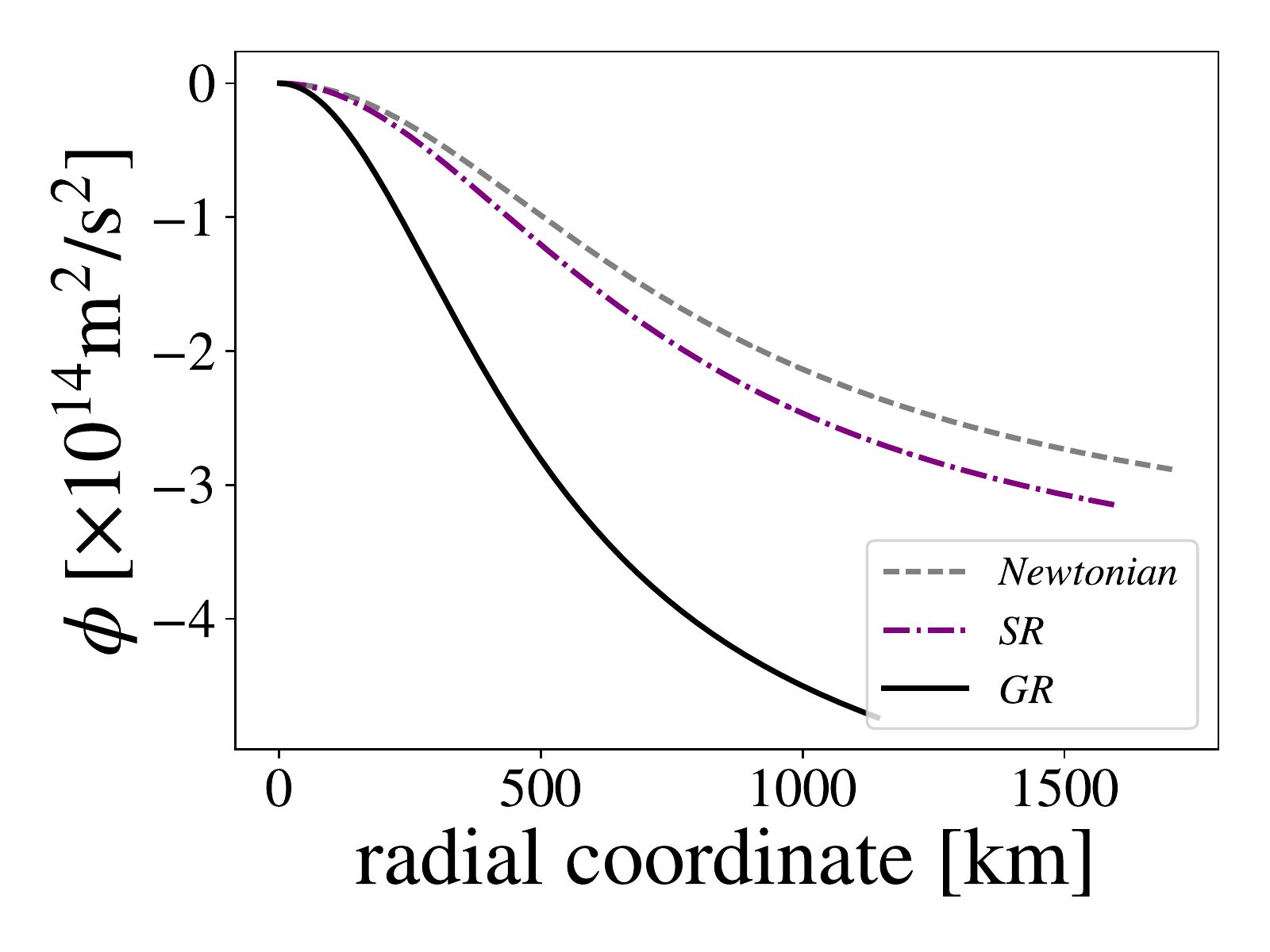}
\caption{General relativistic and Newtonian gravitational potentials as a function of radial coordinate for a fixed total star mass of $1.415M_{\odot}$.}
\label{REALpotentials}
\end{center}
\end{figure}

Fig.\eqref{potentials} displays the gravitational fields of the stars with total mass $1.415M_{\odot}$ as a function of radial coordinate, we can note that the gravitational fields are initially very different, however, outside the star the fields match each other, thus implying that the gravitational field outside the star can be regarded as Newtonian. In addition, inside the star, where the gravitational fields are very different, it can be observed that there is a deviation of about $\sim 200$\% between the correspondent highest values of the gravitational field (the minima of the curves in Fig.\eqref{potentials}), this is due to the very different central densities of the three cases (see Tab.\eqref{tabeladensidades} and Fig.(\ref{varios}b)). In particular, the central density $\rho^{{\rm GR}}_C$ is about $4$ times larger than $\rho_C^{{\rm Newton}}$, for the fixed total mass of $1.415M_{\odot}$. We also display in Fig.\eqref{REALpotentials} the gravitational potentials of WD with total mass $M=1.415M_{\odot}$, from we can note a fairly difference between them. 

\begin{figure}[h]
\begin{center}
\includegraphics[width=0.8\linewidth]{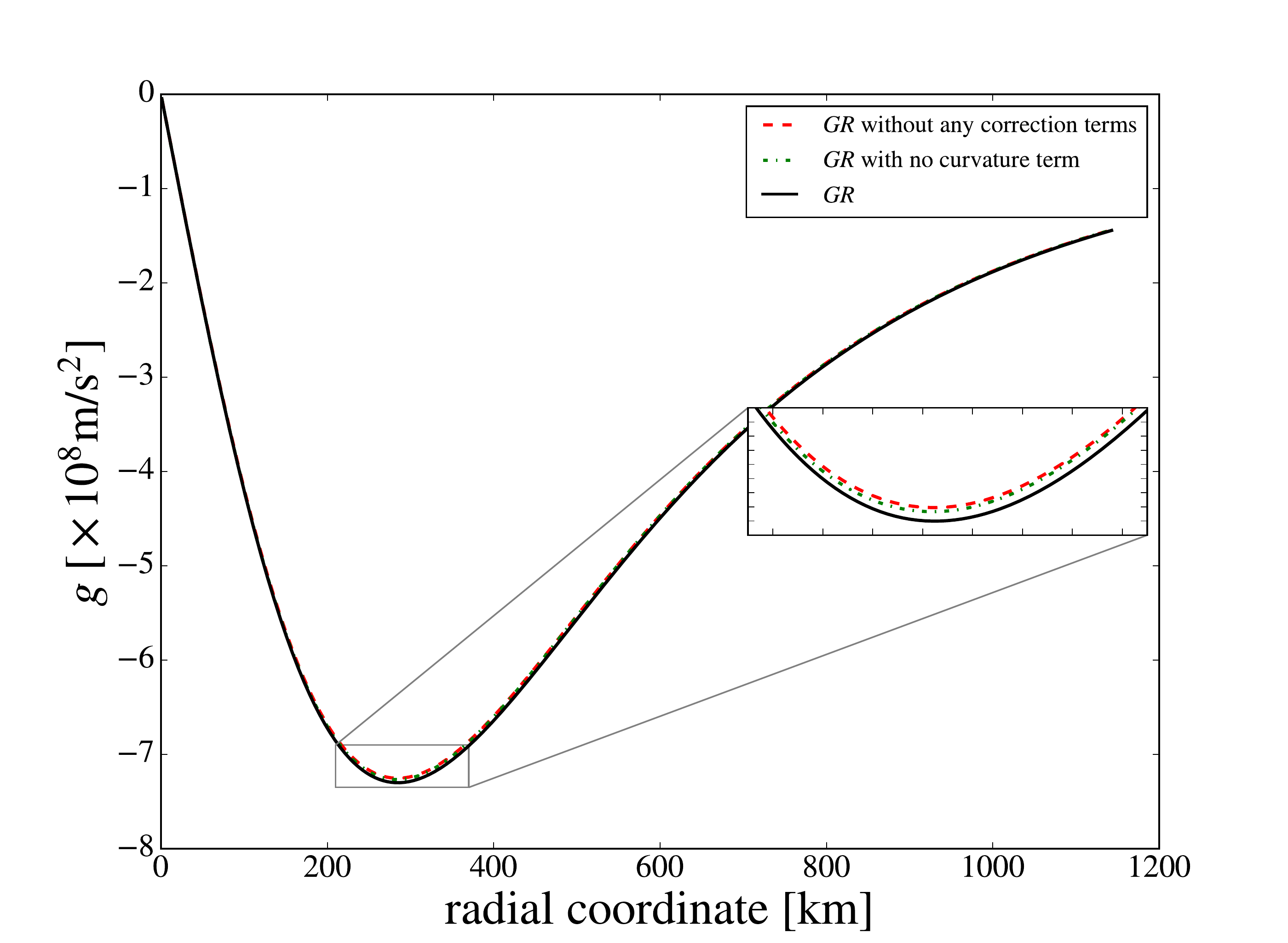}
\caption{General relativistic gravitational fields as a function of radial coordinate for a total mass of $1.415M_{\odot}$, calculated with and without correction terms.}
\label{gravfields}
\end{center}
\end{figure}

In Fig.\eqref{gravfields} we show for the same mass of $1.415M_{\odot}$ the general relativistic gravitational field, calculated in three different ways: with all corrections terms, with no curvature term and without any correction term. {\it A priori}, from the Fig.\eqref{gravfields} the correction terms seems to be not relevant as well, however, they yields to the important effect observed in the case of fixed total masses, thus allowing larger densities near to the center of the star $r<300$km (see Fig.(\ref{varios}b)).

\subsubsection{Radius relative difference}\label{Rdiff}
Henceforth, in the present work we compare just the results of the purely Newtonian case (i.e., without special relativistic corrections) with the general relativistic outcomes for the equilibrium configurations of WD. For this purpose, we firstly define the quantity 
\begin{equation}\label{diffraio}
\Delta R=\frac{R_{\textrm{Newton}}-R_{\textrm{GR}}}{R_{\textrm{GR}}},
\end{equation}
being $R_{\textrm{Newton}}$ and $R_{\textrm{GR}}$ the radii given, respectively, by the Newtonian and general relativistic cases. Therefore, $\Delta R$ means the relative difference between the values of radii for fixed total masses.  
\begin{figure}[h!]
\begin{center}
\includegraphics[width=0.8\linewidth]{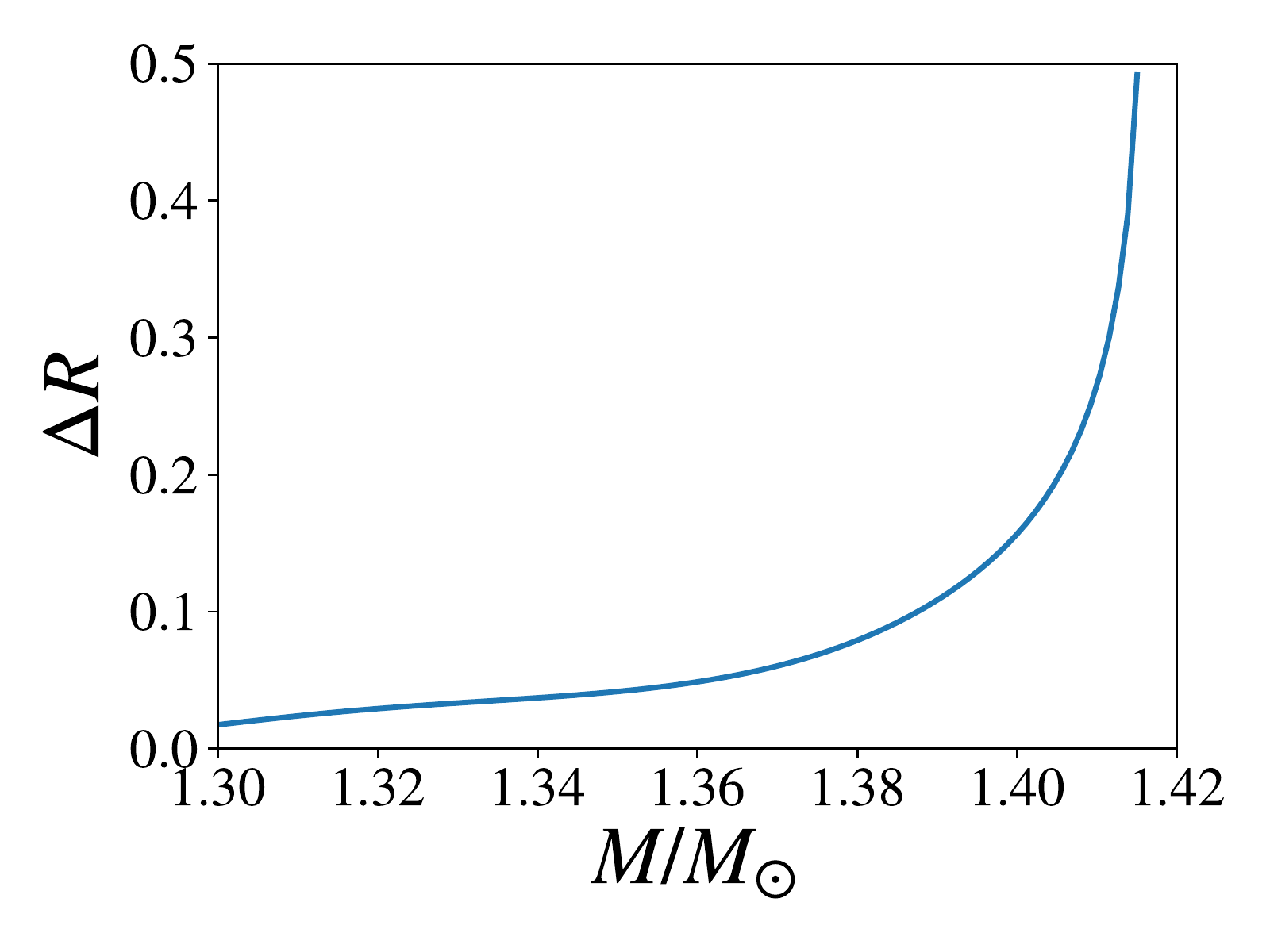}
\caption{Radius relative difference $\Delta R$ versus fixed total star mass.}
\label{plotdeltaR}
\end{center}
\end{figure}

In Fig.\eqref{plotdeltaR} we plot the quantity $\Delta R$ for some values of fixed total mass between $1.3-1.415M_{\odot}$. We can see that this quantity increases very fast when approaching $1.41M_{\odot}$. In fact, when the mass is about $1.41M_{\odot}$ we can see that the relative difference in radius is nearly $50\%$. It is worth to mention that the relative difference in radius is about $37\%$ for a mass of exactly $1.41M_{\odot}$, i.e., the measured mass of the white dwarf $EUVE~J~1659+440$ \cite{Vennes1997}.

\subsubsection{Surface Gravity}\label{surfacegravity}
It is worth to study the general relativistic effects on the surface gravity of the stars since this quantity can be observationally found. 
We calculate the surface gravity $g$ in a Newtonian framework as
\begin{equation}\label{sg1}
g_{\textrm{Newton}}=-\frac{GM}{R^2}.
\end{equation}
To calculate the general relativistic surface gravity we use the expression given by \cite{adler,Ji2008}
\begin{equation}\label{sg2}
g=-\left(\frac{GM}{R^2}\right)\frac{1}{1-\frac{2GM}{c^2R}},
\end{equation}
in which is merely Eq.\eqref{surfgravityGR} for the case of zero pressure.

Using the values of Tab.\eqref{tabelaraios} we calculate the Newtonian surface gravity and general relativistic surface gravity.
\begin{figure}[h]
\begin{center}
\includegraphics[width=0.8\linewidth]{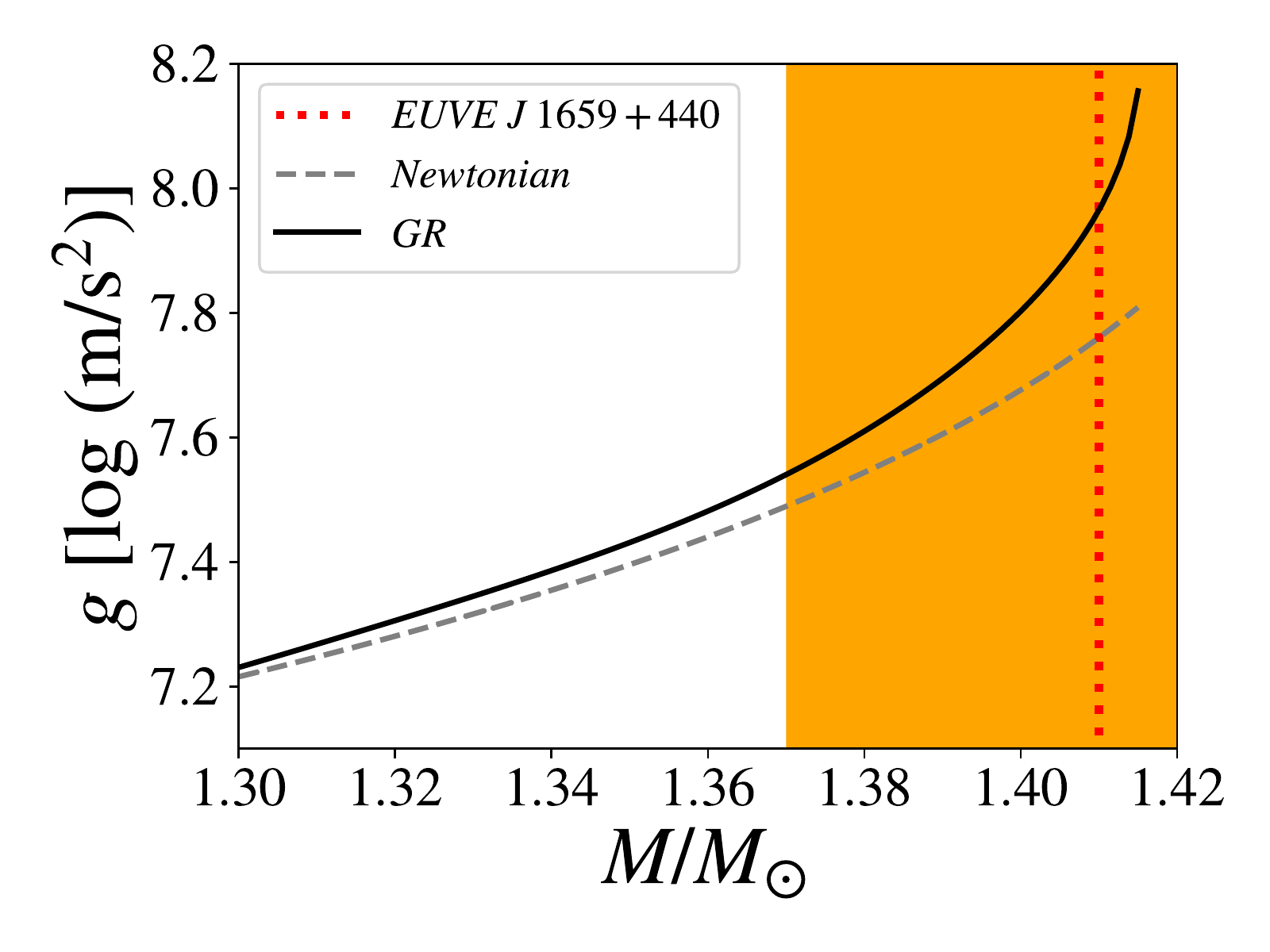}
\caption{Surface gravity versus fixed total star mass. The dotted red line is the measurement of mass of the most massive white dwarf ($EUVE~J~1659+440$) found in literature \cite{Vennes1997} and the shaded orange region is its estimated error.}
\label{gxmassa}
\end{center}
\end{figure}
In Fig.\eqref{gxmassa} is plotted the surface gravity against fixed total masses together with the observational data of the white dwarf $EUVE~J~1659+440$. It is easy to see that using Newtonian results we are sub-estimating the surface gravity of the stars in comparison with general relativistic outcomes.
\subsubsection{Surface gravity relative difference}
Since the values for general relativistic surface gravity are much higher than Newtonian ones we define the quantity
\begin{equation}
\Delta g=\frac{g_{\textrm{Newton}}-g_{\textrm{GR}}}{g_{\textrm{GR}}}.
\end{equation}
\begin{figure}[h]
\begin{center}
\includegraphics[width=0.8\linewidth]{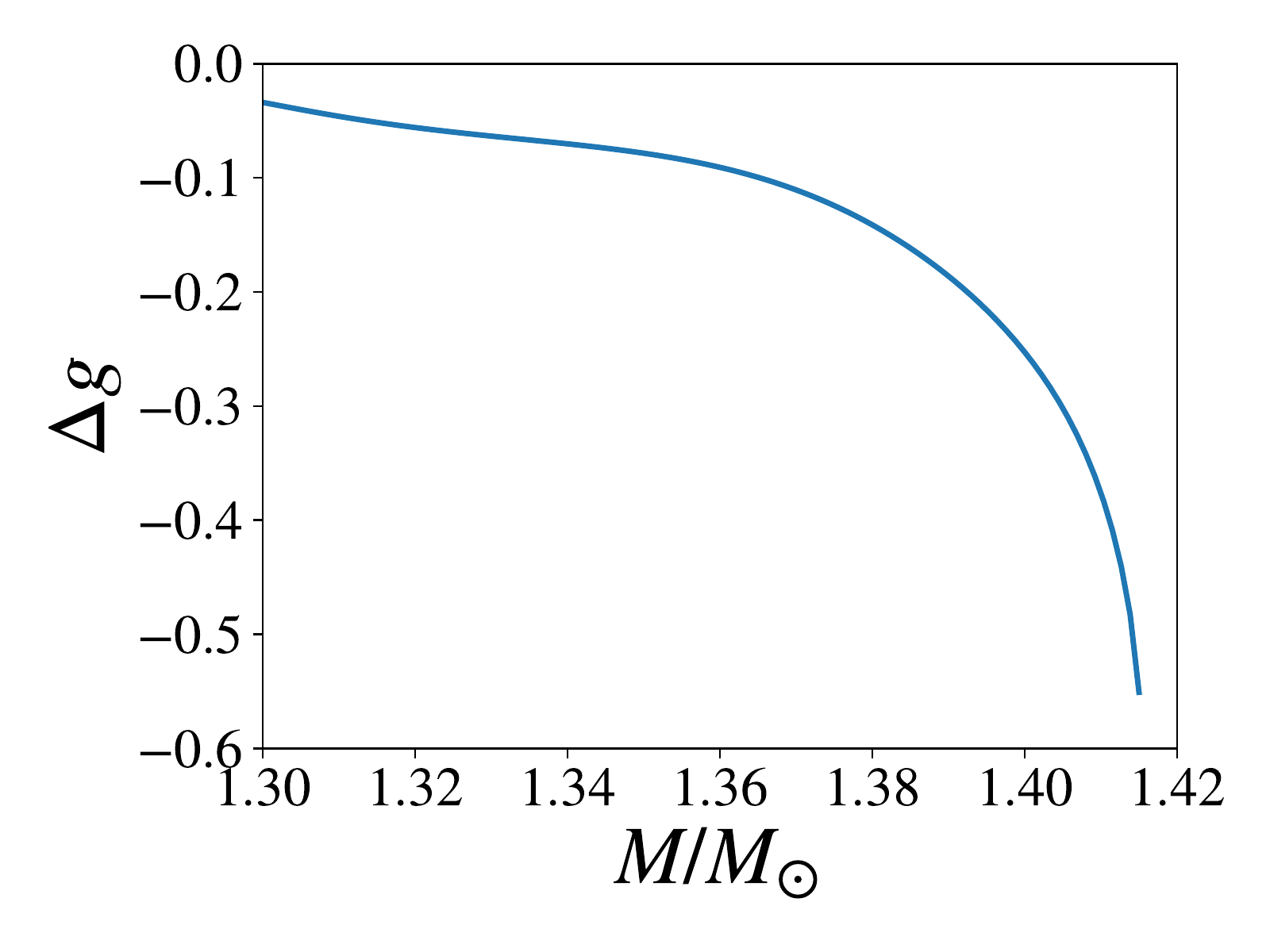}
\caption{Relative difference between Newtonian surface gravity and general relativistic surface gravity against fixed total star mass.}
\label{deltagxmassa}
\end{center}
\end{figure}

The relative difference $\Delta g$ is shown in Fig.\eqref{deltagxmassa}. It is interesting to note that in Fig.\eqref{deltagxmassa}, for a mass of $1.415M_{\odot}$, we have about 55\% of relative difference for the values of surface gravity.

\section{Fit of the general relativistic mass-radius relation}
Keeping in mind the importance of GR for WD, we fit the general relativistic mass-radius relation in order to obtain an analytic expression that better estimate the WD radius, rather than the non-relativistic Newtonian expression
\begin{eqnarray}
\frac{M}{M_{\odot}}=2.08\times 10^{-6} \left(\frac{R}{R_{\odot}}\right)^{-3},
\end{eqnarray}
where $R_{\odot}$ is the radius of the Sun.

The expression we use to fit the general relativistic curve in Fig.\eqref{mrgeral.} is given by
\begin{eqnarray}\label{eqfit}
\frac{M}{M_{\odot}}=\frac{R}{a+bR+cR^2+dR^3+kR^4},
\end{eqnarray}
where $k$ is the inverse of the constant in the non-relativistic Newtonian mass-radius relation $k=(2.08\times 10^{-6}R_{\odot}^3)^{-1}$.

\begin{figure}[h]
\begin{center}
\includegraphics[width=0.8\linewidth]{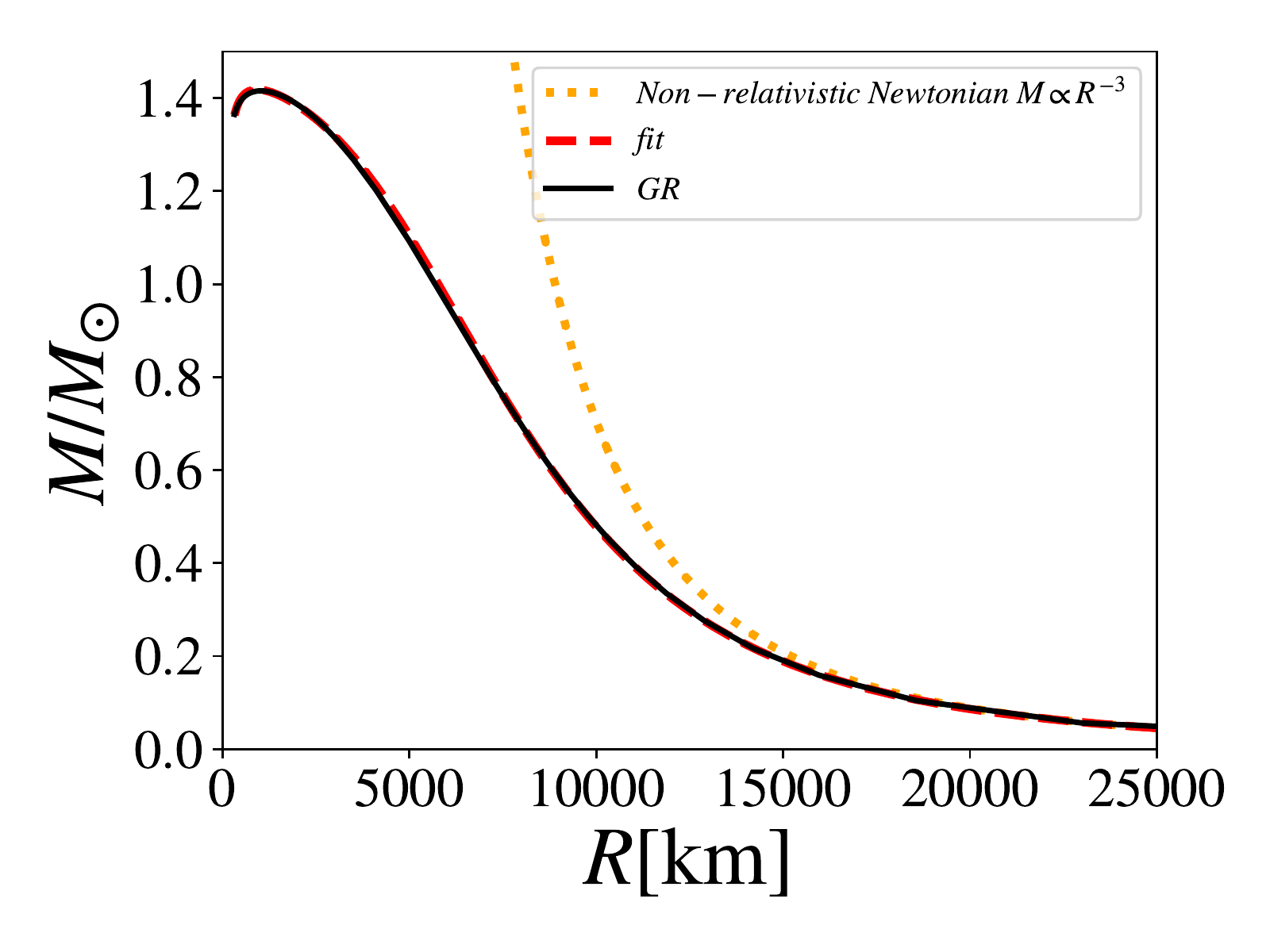}
\caption{Fit of the general relativistic mass-radius diagram with Eq.\eqref{eqfit} (red dashed line) and the non-relativistic limit (dotted orange line).} 
\label{mrfit2}
\end{center}
\end{figure}

The constants $a$, $b$, $c$ and $d$ are parameters that depend on the interior fluid EoS of the star, such that, using the EoS described by Eq.\eqref{eos:simp} and $\mu_e=2$ (Fig.\eqref{mrfit2}) we find
\begin{eqnarray}
a&=&20.86~ \textrm{km} \nonumber\\
b&=&0.66 \nonumber\\
c&=&2.48 \times 10^{-5}~ \textrm{km$^{-1}$}\nonumber\\
d&=&2.43\times 10^{-9}~ \textrm{km$^{-2}$}.
\end{eqnarray}
\begin{table*}[t]
          \caption{Values of the constants for the analytic mass-radius relations.}\label{tabela2}
		\begin{center}
			\begin{tabular}{l c c c c}
				\hline
				Model  & a [km] & b & c $\left[{\textrm km}^{-1}\right]$ & d  $\left[{\textrm km}^{-2}\right]$\\
				\hline
				$\mu_e=2$& $20.86$  & $0.66$ & $2.48\times 10^{-5}$ & $2.43\times 10^{-9}$\\
			        $\mu_e=2.154$	& 15.05	& 0.79 & $3.56 \times 10^{-6}$ & $4.9 \times 10^{-9}$\\
				He& $18.95$ & $0.68$ & $1.84 \times 10^{-5}$ & $-9.85 \times 10^{-10}$\\
				C & $0.79$ & $0.69$ & $1.22 \times 10^{-5}$ & $6.7 \times 10^{-12}$\\
                                O & $-27.06$ & $0.76$ & $-1.21 \times 10^{-5}$ & $3.1 \times 10^{-9}$\\
				\hline
			\end{tabular}
		\end{center}
\end{table*}

We employed Eq.\eqref{eqfit} to depict analytically other mass-radius relations derived from a few EoS models, such as, the Salpeter EoS (for He, C and O stars) and $\mu_e=2.154$. The values of fitted parameters are given in Tab.\eqref{tabela2}.

\section{Concluding Remarks}

In this paper we showed that General Relativity is very important to estimate correctly the radius of a massive WD ($M> 1.3M_{\odot}$) and, consequently, to calculate the surface gravity. We also showed that the minimum radii are very different within either Newtonian or general relativistic cases (about 200\% at most).

We demonstrate that for fixed values of total mass there is a large deviation from Newtonian WD radius to general relativistic WD radius, for example, for a mass close to the value $M= 1.42M_{\odot}$ the Newtonian radius is about 50\% larger than the general relativistic one. For the most massive WD found in literature $M=1.41\pm 0.04$ \cite{Vennes1997} the Newtonian value of radius is now $37$\% larger than the general relativistic one (or at least $6$\% for a mass of $1.37M_{\odot}$). Due to those deviations in radius the surface gravity is expected to be 55\% smaller in Newtonian case in comparison with the result from GR for a fixed total mass of about $M= 1.42M_{\odot}$.

Briefly, the GR effects produces a different correlation between surface gravity and radius, what may induce changes in the values of observational parameters. In particular, if we measured the surface gravity for a massive WD that we know the mass, the correct radius obtained by GR is going to be smaller than the one we would obtain if we do not take into account general relativity, because of the different mass-radius relation of the two cases. 

The WD structure in a general relativistic, finite temperature case was studied in \cite{DeCarvalho2014a}, in which was showed that the finite temperature effects are more significant the less massive the star is. The deviations arising from thermal effects are negligible for stars with $M<1.2M_{\odot}$. On the other hand the main effects of GR appears for stars with $M>1.3M_{\odot}$, what turns both effects crucial for the determination of the WD mass-radius relation from observations. 

We also found a novel analytic mass-radius relation by fitting the general relativistic mass-radius relationship obtained numerically. We suggest that it can be useful to calculate other properties of the stars like magnetic dipole field, moment of inertia, gravitational red-shift and so on.

\begin{acknowledgements}
GAC thanks Coordena\c{c}\~{a}o de Aperfei\c{c}oamento de Pessoal de N\'{i}vel Superior (CAPES) for the financial support. The authors also acknowledge FAPESP for support under the thematic project \# 2013/26258-4.
\end{acknowledgements}

\bibliographystyle{spphys}       
\bibliography{sample}   
%
%

\end{document}